# Realization of a Spin Glass in a two-dimensional van der Waals material

Banabir Pal[1,2,*], Ajesh K. Gopi[1,2], Yicheng Guan[1,2], Anirban Chakraborty[1], Kajal Tiwari[1], Anagha Mathew[1], Abhay K. Srivastava[1], Wenjie Zhang[1], Binoy K. Hazra[1], Holger Meyerheim[1], Stuart S. P. Parkin[1,*]

[1] Max Planck Institute of Microstructure Physics, Halle (Saale) 06120, Germany
[2] Equal author contributions

* email: stuart.parkin@mpi-halle.mpg.de, banabir.pal@mpi-halle.mpg.de

**ABSTRACT**

**Recent advances in van der Waals (vdW) materials have sparked renewed interest in the impact of dimensionality on magnetic phase transitions. While ordered magnetic phases have been demonstrated to survive in the two-dimensional (2D) limit, the quest for a spin-glass with quenched magnetic disorder in lower dimensions has proven elusive. Here we show evidence of a spin-glass emerging from randomly distributed Fe atoms in $Fe_3GeTe_2$, the first time such a state has been reported in a vdW material. AC magnetic susceptibility displays a strong frequency dependence indicative of slow spin dynamics. Additional distinctive phenomena, including ageing, chaos, and memory effects, further substantiate the existence of a glassy state. Remarkably, we find that this state persists even in single-unit-cell thick $Fe_3GeTe_2$, thereby confirming the existence of a 2D spin-glass. The formation of spin-glass states via intercalation in vdW systems allows for highly tunable spin-glass states that are otherwise difficult to realize.**

The critical dimension above which a phase transition can occur has long been of interest (*1*). In particular, magnetism in the 2D limit has proved challenging as theory predicts that, at a finite temperature, long-range magnetic order for Ising and Heisenberg spins only exists in dimensions above one and two, respectively (*2-5*). Recent experimental findings have, however, demonstrated that magneto-crystalline anisotropy (*6-8*), dipolar interactions (*9*) and interfacial spin-orbit coupling (*10*) can give rise to stable ferro/antiferromagnetic (FM/AFM) ordering even in the 2D limit, illustrating the limitations of the Mermin-Wagner theorem (*3*). Since the first example of robust ferromagnetic order in exfoliated 2D vdW materials, a large number of magnetic ground states including ferro (*6-8, 11-18*), 2D-XY (*19*), antiferro (*20-22*) and helical (*23, 24*) magnets have been unveiled, which have, thereby, significantly expanded the phase space of ordered 2D magnets (*1, 25*).

Conspicuously missing from this list is the spin glass state (*26*), a quench-disordered magnetic phase characterized by frustrated spin interactions that prevent long-range magnetic order. The role of critical dimensions in spin-glass phase transition remains multifaceted and is yet to be fully understood. Investigations into the 2D random Ising model (*27-29*) suggest no phase transition at nonzero temperatures, while highlighting an algebraic divergence of the correlation length at zero temperature, hinting at a possible 'zero temperature' spin-glass phase transition (*30*). For Ising systems, the lower critical dimension of this 'zero temperature' phase transition appears to lie at or above two dimensions, depending on the disorder (bimodal or Gaussian) distribution (*31*). For XY and Heisenberg spin-glasses with short-range interaction, three dimensions falls below the lower critical dimension (*32-34*). Thus, from a theoretical standpoint, stabilizing a 2D spin glass is difficult. Another challenge is the creation of the 'site-disorder' within a 2D vdW material. The spin glass phase was initially realized in a site-disordered, dilute magnetic lattice with RKKY-like exchange interactions (*35, 36*). Replicating this in a 2D vdW material is difficult, requiring either naturally occurring magnetic site vacancies or vacancies induced through intercalation. Therefore, there exists several open questions; in particular, can a spin glass phase exist within a bulk 2D vdW material? If so, would it persist in the 2D limit, and how would its behavior evolve as the material's thickness decreases?

In this work, we provide evidence for a spin glass in a vdW magnet, namely in single-crystalline flakes of $Fe_3GeTe_2$ (FGT), the first time such a state has been reported in a vdW material. The temperature dependence of the AC magnetic susceptibility shows a strong frequency dependence characteristic of a glassy state with slow spin dynamics. We also demonstrate other specific phenomena, namely 'ageing', 'chaos', and 'memory' effects, which are hallmarks of a spin glass ground state. Having demonstrated the existence of a quenched disorder spin state in bulk FGT, we investigate the evolution of such a ground state as we reduce the thickness and approach the 2D limit. Remarkably, we find that the glassy dynamics survives even for single unit cell thick FGT flakes, thereby providing a new platform to understand experimentally the response of the spin glass in the 2D limit. The robustness of this state in the 2D limit also prompts questions about its stability mechanisms.

Here we take advantage of structural defects that often give rise to interesting properties in 2D vdW materials. In particular, it was recently shown that $Fe_3GeTe_2$ exhibits several distinct types of structural defects (*37-39*), including, incomplete occupancy of specific Wyckoff sites, and interstitial Fe atoms within the vdW gap (*40*). We carried out an X-ray crystallographic analysis of the FGT crystals using a high brilliance Ga-jet x-ray source and a six-circle diffractometer that is capable of probing the detailed crystal structure of even ultra-thin flakes and microcrystals. About 40 symmetry independent reflections (point symmetry group 3m) were collected from which the structure factor magnitudes $|F_{obs}(HKL)|$ were derived. Defects from the nominal structure reduce the crystal symmetry to the space group (SGR) P3m1 (Nr. 156). The high symmetry of this SGR means that only the z-parameters of the atoms need to be varied, in addition to the Debye-Waller factors, an overall scale factor and the atomic site occupancy factors (SOF). The best fit of the calculated structure factor magnitudes, $|F_{calc}(HKL)|$, to the observed ones is achieved by introducing a fraction of 0.2 to 0.3 of Fe vacancies into the Wyckoff sites (1b and 1c site in SGR 156) next to the Ge atoms (see Supplementary materials (SM) for details). Moreover, we find a random distribution of Fe atoms within the vdW gaps, with a concentration of ~10-15%, resembling a dilute magnetic lattice. Although this is a quite subtle structural modification ($\approx$ 8 electrons) in comparison with the total charge density within the unit cell (in total 428 electrons per unit cell for stoichiometric FGT), our highly precise data allows the direct and unambiguous identification of these fractionally occupied sites. Fig. 1A shows the model of the as-determined FGT structure and Fig. 1B shows the calculated charge density $\rho(x, y, z)$ projected along [0001] from the Fourier-Transform (FT) of the $F_{obs}$. It is dominated by the contribution of the atoms at the (1b) and (1c) sites (Te, Ge, Fe) which form a chain running along the [0001] direction. Some smaller density is also observed at the (1a) site at (000) related to Fe atoms. Fig. 1C shows the difference in charge density, $\rho\Delta(x,y) = \rho_{obs} - \rho_{calc}$, which is the FT of $F_{obs} - F_{calc}$. On a scale enhanced by a factor of 10 relative to $\rho(x, y)$, it exhibits a clear positive maximum at the site 1a, indicating the presence of charge density at this site. The difference in density vanishes if Fe with a site occupancy of 0.15 is placed into the two vdW sites of the unit cell. In addition, the fit quality improves by 15% (see SM for details). Thus, we establish that there is a randomly occupied 2D array of Fe sites within the vdW gap which we hypothesize can form a 2D spin-glass.

Next, we carry out field-cooled (FC) and zero field-cooled (ZFC) DC magnetization measurements, which show a clear magnetic phase transition at around 200 K, together with an additional anomaly at a lower temperature (Fig. 1D). Below the transition temperature, the FC and ZFC curves diverge, a phenomenon commonly observed in spin glass systems. This divergence has been previously attributed to thermo-remnant magnetization (*41*). However, it is important to note that low-temperature splitting of the FC/ZFC curves only signifies the onset of magnetic irreversibility, not necessarily the collective behavior associated with a spin glass. Therefore, to further explore the dynamic nature of the phase transition, we carry out frequency-dependent AC susceptibility measurements with a small, out-of-plane magnetic field. Both real ($\chi'$ in Fig. 1E) and imaginary ($\chi''$ in Fig. 1F) parts of the AC susceptibility data show a sharp frequency-dependent dispersion at around 200 K, suggesting the presence of slow spin dynamics (*42, 43*). Notably, the freezing temperatures in $\chi''$ (Fig. 1F) display a strong frequency dependence, a hallmark of a spin glass. On the other hand, the system also displays another minor feature at around 125 K, which is less prominent in $\chi'$, and which occurs at the same temperature in $\chi''$ as a function of frequency. Therefore, we focus our attention on the transition at approximately 200 K, where a substantial frequency-dependent dispersion is observed. Carrying out similar investigations on other 2D magnets, such as the antiferromagnetic 2D $MnPS_3$ (see details in Fig. S4), we rather observe a well-defined antiferromagnetic phase transition. And, in contrast to the FGT system, the FC/ZFC data in $MnPS_3$ coincide below the phase transition temperature, and the AC susceptibility measurements show no frequency dependence near the transition temperature. These findings imply that FGT exhibits characteristic slow spin dynamics indicative of a spin-glass transition, while $MnPS_3$ undergoes an antiferromagnetic transition without any spin glass behavior.

The spin glass is an out-of-equilibrium state whose response to an external field is logarithmically slow at low temperatures (*26*). We investigate this 'ageing' phenomenon in the FGT crystals on the scale of several hours by measuring the temporal evolution of $\chi''$ (Fig. 2A) using a small AC magnetic field (~ 0.5 Oe) and at different low frequencies ($\omega/2\pi = 0.1$ to 10 Hz). Consistent with earlier observations (*41*), we find that the relaxation magnitude of $\chi''$ increases as the frequency decreases, which is a characteristic behavior of spin glass systems. We also find that in the infinite time limit (here after six hours) $\chi''$ remains non-zero and converges to a finite limit. As is expected for an 'ageing' process, $\chi''$ decreases with time (Fig. 2A) since the system becomes 'stiffer' with increasing age, resulting in a slower response to a changing external field. It is well known that the divergence of the AC susceptibility (both $\chi'$ and $\chi''$ in Fig. 1E and 1F) close to the glass temperature ($T_g$) is associated with the divergence of a spin-spin correlation length, a similar phenomenon also seen in ordered ferro- and antiferromagnetic systems. Therefore, it is tempting to relate the 'ageing' process (Fig. 2A) with the progressive growth of magnetic domains towards the global equilibrium. Nonetheless, this simplistic description fails to encompass other experimental observations, as discussed in the next section, that are characteristic of a spin glass.

A typical spin glass system, along with 'ageing', also exhibits 'chaos', and 'memory' effects which can be understood from a hierarchical distribution of metastable states (*44, 45*), as illustrated schematically in Fig. 2B and 2C. The hierarchical model predicts the existence of a vast multitude of nearly degenerate ground states. As the temperature is reduced, these metastable states further split into larger numbers of new states (Fig. 2C). Therefore, the equilibrium spin configuration of a spin glass at one temperature ($T_1$) is distinct from its equilibrium state at another lower temperature ($T_2 = T_1 - \Delta T$). This picture is in stark contrast with a progressive domain growth model where the global equilibrium configuration is independent of the temperature. As a result of these ultra-metric topologies of the energy landscape associated with a spin-glass (*46*), such systems exhibit a 'chaos' effect, where the $\chi''$ suffers a jump even when the temperature is reduced. We have systematically investigated this effect by probing the time-dependent evolution of $\chi''$ under negative temperature cycling (Fig. 2D). Phase-I of this protocol involved quenching the sample from room temperature to a sub-glass transition temperature ($T_1$) and keeping it at $T_1$ for $t_1 = 1.5$ hr. Following this, in phase-II, we further lowered the temperature to $T_2$, where the sample was again held isothermally for a duration of $t_2 = 2.5$ hr. Finally, in phase-III, the temperature was raised back and maintained at $T_1$ for an additional $t_3 = 2$ hr. Throughout this process, we continuously recorded the relaxation of χ". Similar to Fig. 2A, in phase-I we see an ageing effect where χ" decreases systematically. In phase-II, if the temperature difference $\Delta T$ ($\Delta T \equiv T_1 - T_2$) is small (e.g. $\Delta T = 0.2\ K$ for the green line in Fig. 2D) we observe a similar decay, since we suppose that the change in the energy landscape is negligible. However, if $\Delta T$ is large, then due to hierarchical distributions of the metastable states, the equilibrium spin configuration in $T_1$ will be different from $T_2$. As a result, when the sample is cooled to $T_2$, initially the spins experience a randomization due to the drastic modification of the energy landscape and consequently $\chi''$ exhibits a sudden jump (arrow in blue line in Fig. 2D). This is known as a thermal 'chaos' effect (also called a rejuvenation effect). We also find evidence of memory effects in our FGT system, where the state reached by the system at a given temperature can be retrieved even after a negative temperature cycle. For example, we find that for the case $\Delta T = 2\ K$ (blue in Fig. 2D) the system returns back to the same value for $\Delta T = 0.2\ K$ (green in Fig. 2D). The presence of thermal chaos and memory effects further confirms the emergence of a spin-glass ground state in our FGT crystals.

We also find a similar field induced rejuvenation of the ageing process in field cycling experiments (Fig. 2E). Here, after the initial quench from 300 K to $T_1$ (phase-I) in zero field, the relaxation in $\chi''$ is recorded for a time $t_1$ =2 hr and then in Phase-II, a DC field is applied and the relaxation in $\chi''$ is continuously recorded for a time $t_2 = 3$ hr. In phase-III, the relaxation in $\chi''$ was measured for an additional time $t_3 = 2$ hr after removing the DC field. Consistent with previous reports for conventional spin glasses (*41*), we find that both increasing and decreasing the DC field reinitialize the relaxation process (Fig. 2E). A jump in $\chi''$ in phase-II and phase-III supports this picture. The magnitude of the jump in $\chi''$ (Fig. 2D) is found to be larger when a larger DC field is used in the aforementioned protocol.

In addition, we carried out AC susceptibility experiments at $\omega/2\pi = 0.1$ Hz and $H = 0.3$ Oe with a different protocol to demonstrate the same ageing, chaos and memory effects. Here, first, we cool the sample from above $T_g$ down to ~176 K at a constant rate of 0.1 K/min, and then heat the sample back continuously at the same rate (Fig. 2F). In the second cycle, we start the cooling from a temperature above $T_g$, but wait at an intermediate temperature $T = 186\ K$ for 8 hours. Notably, during this waiting period, we find that $\chi''$ (Fig. 2F) relaxes downwards which again is consistent with the ageing effect. After the ageing period at 186K, the cooling procedure resumes, and we find that $\chi''$ merges back with the reference curve only a few Kelvin below $T_1$. Therefore, a temperature difference of ~2 K renormalizes the energy landscape and $\chi''$ completely loses its ageing effect due to the chaos effect. Furthermore, the memory effect was observed during the second heating process, where $\chi''$ displayed a down turn due to the 8-hour wait period introduced during the second cooling phase. Collectively, these experiments provide compelling evidence of a spin glass-like phase transition.

Upon establishing a spin glass state in a bulk FGT flake, our focus shifts towards exploring whether this ground state persists as we approach the 2D limit. Our first step is to examine how the 'disorder' distribution within the vdW gap is affected by reducing sample thickness. We carry out crystallographic analysis on three different samples of similar lateral size (~ 20 µm) with thicknesses ranging from 20 to 50 nm. We find no significant differences in the structure and the disorder distribution even in the thinnest sample (See SM for details). This demonstrates that defect distribution within the vdW gap does not change significantly down to a thickness of 20 nm.

Since traditional SQUID-based techniques lack the sensitivity to measure spin dynamics in exfoliated flakes, we developed two AC measurement methods utilizing the magneto-optical Kerr effect (MOKE) and the Hall Effect (see SM for details of these techniques and analysis process). Figure 3A and 3B depict the temperature dependence of $\chi'$ at various frequencies for a single FGT Hall bar with a thickness of ~17 nm (inset of Fig. 3B), as measured using Hall transport measurements and MOKE microscopy, respectively. Both the Hall transport and MOKE results reveal a frequency-dependent $\chi'$, confirming the presence of slow spin dynamics in even such a thin flake. The consistency in the results from the two distinct AC susceptibility measurement techniques (Fig. 3A and 3B) is noteworthy. We have also measured anomalous Hall data, which exhibit rectangular hysteresis loops at lower temperature. The remanent $R_{xy}$ and coercive field decrease at higher temperatures, eventually vanishing completely above 185 K.

Spin dynamics in ultrathin FGT flakes were investigated using Magneto-optical Kerr Effect (MOKE) microscopy. Fig. S6 in supplementary material shows the AFM image of one such flake where the minimum thickness was found to be a single unit-cell. First, we carried out magnetic hysteresis measurements on the single unit-cell FGT flake using MOKE (Fig. 3D). We find that the transition temperature of the thinnest flake was around 95 K. Remarkably, even in this single-unit-cell FGT flake, we observed frequency dispersion in the real part of the AC susceptibility ($\chi'$, Fig. 3E). These results support the emergence of a true 2D spin-glass state. To elucidate the thickness dependence, similar experiments were conducted on flakes with varying thicknesses. The resulting phase diagram (Fig. 3F) depicts the spin-glass phase as a function of both thickness and temperature. Interestingly, the transition temperature remains independent of thickness down to 17 nm. Below this thickness, a rapid decrease in transition temperature is observed, reaching around 95 K for the single-unit-cell sample. Importantly, glassy dynamics persist throughout the entire thickness range.

In summary, our studies have revealed a spin glass ground state in the 2D limit which takes advantage of the property of self-intercalation within a van der Waals magnetic system. Numerous properties characteristic of a spin glass are observed including ageing, chaos, and memory effects. We anticipate that our demonstration of a spin glass ground state in a 2D van der Waals systems will pave the way for much research concerning, on the one hand, the robustness of the spin glass state in the 2D limit, and, on the other hand, how universal this behavior is in other self-intercalated van der Waals systems.

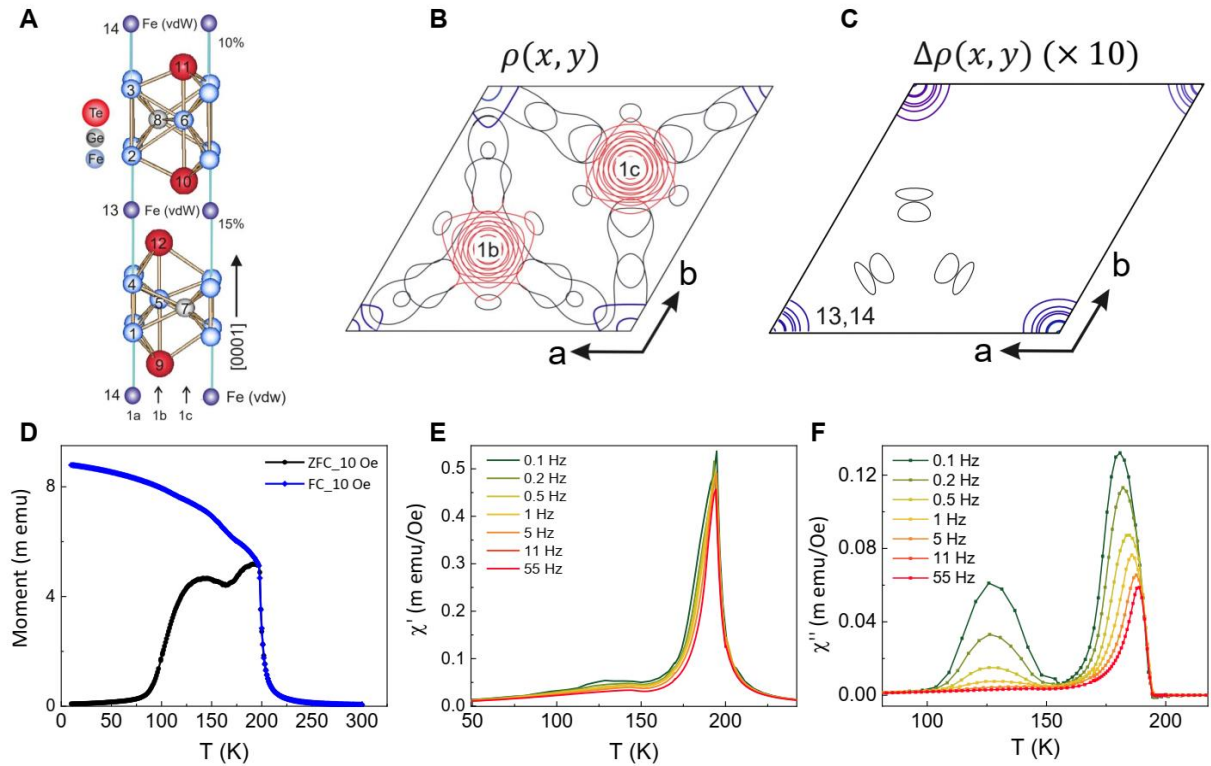

**Fig.1: Structural and magnetic analysis of Fe₃GeTe₂**. (**A**) Structural model of FGT from a side view. Fe (blue), Ge (grey) and Te (red) atoms are arranged in chains along [0001] at Wyckoff sites 1a (00z), 1b ($^1/_3$ $^2/_3$ z) and 1c ($^2/_3$ $^1/_3$ z). vdW site Fe atoms (#13, #14) are shown by purple balls. (**B**) Contour plot of the electron charge density magnitude, $\rho(x,y)$, projected along z along one unit cell. $\rho(x,y)$ is dominated by the Te atoms arranged along [0001] at sites 1b and 1c (red contours). Weak charge density at 1a related to Fe atoms is also observable. Remaining weak densities at asymmetric sites are attributed to truncation effects resulting from the finite number of reflections. (**C**) Difference charge density contour plot, $\Delta\rho(x,y)$ exaggerated by a factor 10 relative to $\rho(x,y)$. $\Delta\rho(x,y)$ is the FT of ($F_{obs}-F_{calc}$) which shows a clear density at 1a (00z) if the vdW Fe sites are not taken into account in the structural model. (**D**) Field-cooled and zero-field-cooled temperature dependences of the magnetization for an applied DC field of ~10 Oe. Temperature dependence of the (**E**) real ($\chi'$) and (**F**) imaginary ($\chi''$) parts of the AC susceptibility measured at an applied field of 0.5 Oe and various frequencies as indicated in the figure. The data is taken on warming after cooling in zero field. All the data in this figure correspond to single crystals of FGT.

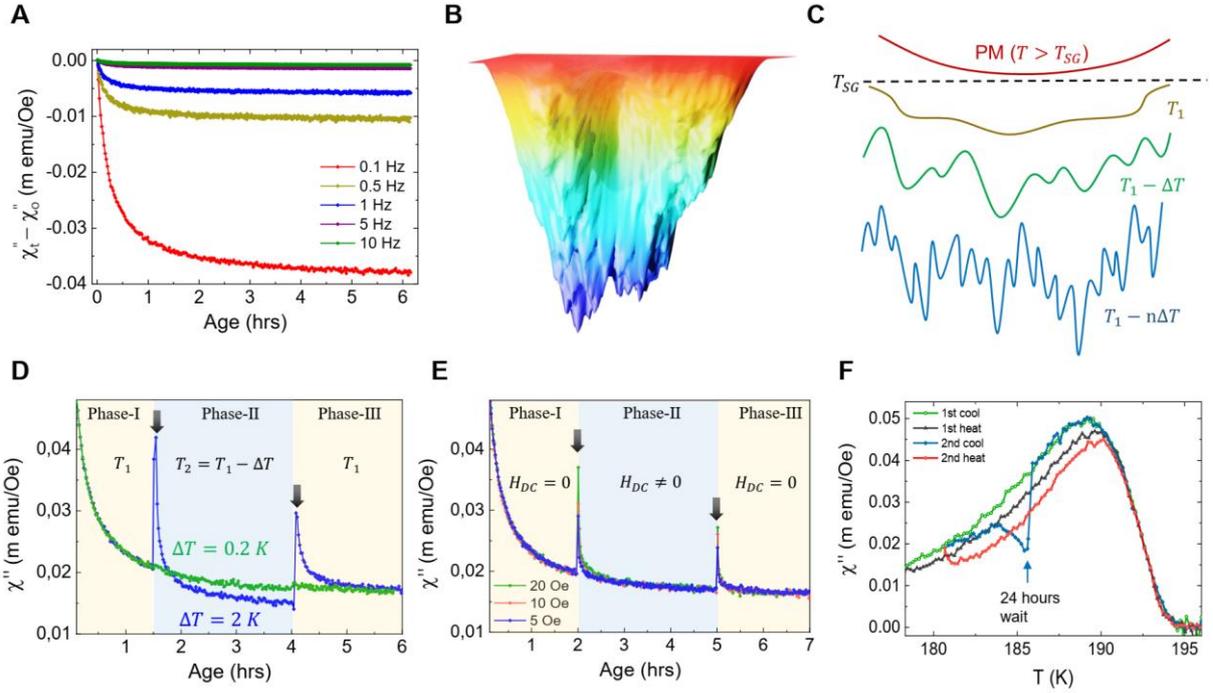

**Fig.2: Ageing, chaos and memory effects in Fe$_3$GeTe$_2$**. (**A**) Evolution of the out-of-phase susceptibility ($\chi''$) over time at different frequencies, illustrating the ageing phenomenon. Each curve is set to zero at time $t = 0$ by subtracting the initial magnitude of $\chi''$ for a better comparison. (**B** and **C**) Schematic depiction of the hierarchical structure of the energy landscape in a spin glass as a function of decreasing temperature (top to bottom) in (B) a 2D plane and (C) along an arbitrary direction. These illustrate the increasing complexity of the system as temperature decreases. (**D**) Variation of $\chi''$ throughout a negative temperature cycle where the temperature is reduced from $T_1 = 186$ K by $\Delta T=0.2$ K (green curve) and $\Delta T=2$ K (blue curve) at t= 1.5 hrs and returned to $T_1$ at $t = 4$ hrs. The DC field is always set to zero. These data show ageing phenomenon in phase-I (T1), rejuvenation/chaos in phase-II (T2), and memory effects in phase-III (T1). (**E**) Variation of $\chi''$ during a similar experiment showing instead a field induced chaos/rejuvenation effect when a DC field is applied at $t = 2$ hrs and set back to zero at $t = 5$ hrs. (**F**) Chaos and memory effects during cooling and heating cycles (DC field = 0). All the data in this figure correspond to single crystals of FGT.

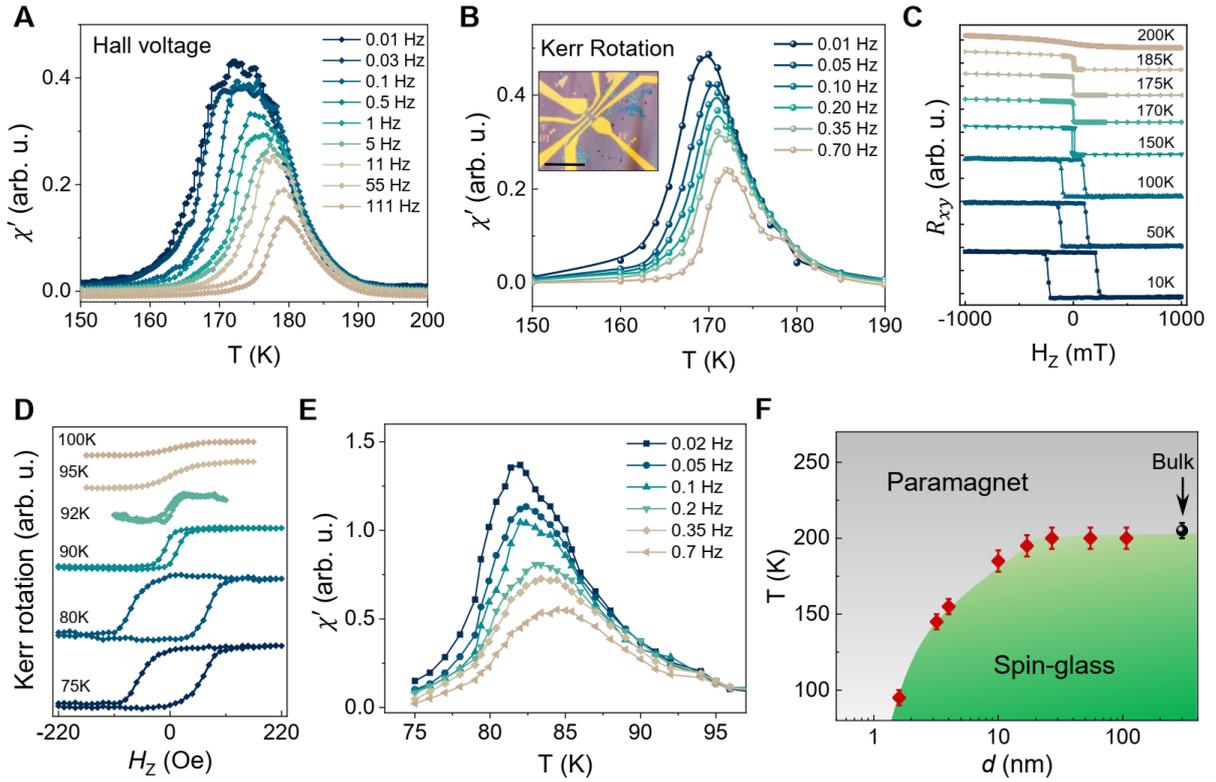

**Fig. 3: Spin glass state in Fe$_3$GeTe$_2$ in the two-dimensional limit**. (**A** and **B**) Real part of the AC susceptibility ($\chi'$) data obtained from (A) Hall transport and (B) MOKE microscopy techniques for the sample shown in in the inset of Fig. 3(B). The sample is an exfoliated flake that is 17 nm thick. Electrical contacts are fabricated onto the flake using Ti/ Au bilayers. The scale bar corresponds to 50 μm. Both these datasets show frequency dependent dispersion in $\chi'$ indicating a slow spin dynamics. (**C**) Anomalous Hall data of the same sample (inset of Fig. 3B) at different temperatures, as indicated in the figure. Both the coercive field and remanent magnetization decreases as the temperature approaches the Curie temperature (Tc ~190K). (**D**) Anomalous Hall effect of the thinnest sample shows a Curie temperature of ~95 K. (**E**) Real part of the AC susceptibility ($\chi'$) data for the thinnest sample shows that slow spin-dynamics persist even in the 2D limit. (**F**) Temperature-thickness magnetic phase diagram for the Fe$_3$GeTe$_2$ system.